\RequirePackage{fixltx2e}
\documentclass[aps,prb,reprint,floatfix, superscriptaddress]{revtex4-2}
\usepackage{latexsym}
\usepackage{amsmath}
\usepackage{amsfonts}
\usepackage{amssymb}
\usepackage{amsthm}
\usepackage{physics}
\usepackage{qcircuit}
\usepackage{graphicx}
\usepackage{dcolumn}
\usepackage{bm}
\usepackage{newfloat,algcompatible}
\usepackage{algorithm}
\usepackage{algpseudocode}
\usepackage{etoolbox}
\usepackage[table]{xcolor}
\usepackage{multirow}
\usepackage{tablefootnote}
\usepackage{booktabs}
\usepackage[caption=false]{subfig}

\DeclareMathOperator*{\argmin}{arg\,min}

\begin{document}
\title{
Growth reduction of similarity transformed electronic Hamiltonians in qubit space
}

\author{Robert A. Lang}
\affiliation{Department of Physical and Environmental Sciences,
  University of Toronto Scarborough, Toronto, Ontario, M1C 1A4,
  Canada}
\affiliation{Chemical Physics Theory Group, Department of Chemistry,
  University of Toronto, Toronto, Ontario, M5S 3H6, Canada}
  
\author{Aadithya Ganeshram} 
\affiliation{Department of Physical and Environmental Sciences,
  University of Toronto Scarborough, Toronto, Ontario, M1C 1A4,
  Canada}
\affiliation{Chemical Physics Theory Group, Department of Chemistry,
  University of Toronto, Toronto, Ontario, M5S 3H6, Canada} 
\author{Artur F. Izmaylov}
\email{artur.izmaylov@utoronto.ca}
\affiliation{Department of Physical and Environmental Sciences,
  University of Toronto Scarborough, Toronto, Ontario, M1C 1A4,
  Canada}
\affiliation{Chemical Physics Theory Group, Department of Chemistry,
  University of Toronto, Toronto, Ontario, M5S 3H6, Canada}

\date{\today}

\begin{abstract}
Accurately solving the electronic structure problem through the variational quantum eigensolver (VQE) is hindered by the available quantum resources of current and near-term devices. One approach to relieving the circuit depth requirements for VQE is to ``pre-process" the electronic Hamiltonian by a similarity transformation incorporating some degree of electronic correlation, with the remaining correlation left to be addressed by the circuit ansatz. This often comes at the price of a substantial increase in the number of terms to measure in the similarity transformed Hamiltonian. In this work, we propose an efficient approach to sampling elements from the complete Pauli group for $N$ qubits which minimize the onset of new terms in the transformed Hamiltonian, while facilitating substantial energy lowering. { We benchmark the growth-mitigating generator selection technique for ground state energy estimations applied to models of the H$_4$, N$_2$ and H$_2$O molecular systems. It is found that utilizing a selection procedure which obtains the growth-minimizing generator from the set of operators with maximal energy gradient is the most competitive approach to reducing the onset of Hamiltonian terms while achieving systematic energy lowering of the reference state.}
\end{abstract}

\maketitle

\section{Introduction}

A broadly investigated application of the variational quantum eigensolver (VQE) is accurately solving instances of the electronic structure problem. \cite{qchem_review2019, qchem_review2020, vqe_review} While the state-of-the-art in digital quantum computing hardware is rapidly improving,  \cite{hardware_review2021} the application of widely investigated fixed circuit ans\"atze, such as the unitary coupled cluster with singles and doubles (UCCSD), \cite{Peruzzo2014, Hempel2018, Romero2018, Lee:2019/jctc/311, Sokolov2019, Mizukami2019, UCCSD_review2022} can require circuit depths beyond the capability of current and near-term devices. To explore the possibilities of near-term quantum advantage, adaptive schemes within the VQE framework featuring system-specific iterative ansatz construction have been proposed. \cite{ADAPT, QCC, qADAPT, iQCC, qeb_ansatz, Zhang_2021, adapt_w_rdms} These approaches aim to adaptively parameterize the target electronic wavefunction to desirable accuracy with a quantum circuit as shallow as possible. Such approaches feature selection of a unitary generator from a predefined operator set (``pool"). At each iteration, the operator pool elements are ranked by  an importance measure. The importance measure has typically been taken to be the absolute value of the energy's first partial derivative with respect to the generator's free parameter. The element corresponding to the maximum value of this importance measure is then taken as generator of the current step unitary included in the quantum circuit ansatz. 

While adaptive schemes have demonstrated robust numerical convergences to ground and excited electronic states to high accuracy, \cite{ADAPT, iQCC, qeb_ansatz, ADAFT} certain instances of the electronic structure problem may demand many iterations within an adaptive VQE scheme, leading to regrettably deep quantum circuits. One approach to reduce the depths of the required circuits is via preprocessing of the electronic Hamiltonian by an exact or approximate similarity transformation of the electronic Hamiltonian. Such transformations aim to admit a more compact circuit description of the target electronic state. In the case of exact transformation, an effective Hamiltonian is obtained as
\begin{align}
    \hat H_{\rm{eff}} = \hat T^{-1} \hat H \hat T.
\end{align}
Approximations to $\hat H_{\rm{eff}}$ can be obtained by neglecting higher order terms in the Baker-Campbell-Hausdorff (BCH) expansion of $\hat T^{-1} \hat H \hat T$, for instance. It is desirable to utilize $\hat T$ which admits a tractable exact transformation of $\hat H$, as in this case $\hat H_{\rm{eff}}$ and $\hat H$ are formally isospectral. For Hamiltonian preprocessing prior to a variational treatment, it is also desirable that the transformation is unitary, preserving the Hermitian property of the Hamiltonian. Hence, exact unitary transformation of $\hat H$ ensures that a variational solution to $\hat H_{\rm{eff}}$ is lower bounded by the ground state energy of $\hat H$. Nevertheless, often motivated by computational feasibility, many recent contributions have considered preprocessing the electronic Hamiltonian with approximate transformations. \cite{DUCCFormalism, Metcalf2020, Motta2020, ADAFT, Evangelista2022} 

An adaptive method employing exact and unitary Hamiltonian transformations formulated in the qubit space algebra is the iterative qubit coupled cluster (iQCC) approach.\cite{iQCC} The iQCC method features selection of individual Pauli products, i.e., tensor products of the Pauli matrices and the single qubit identity, 
\begin{align} 
\hat P_\alpha = \hat \sigma_1^{(\alpha)} \otimes \hat \sigma_2^{(\alpha)} \hdots \otimes \hat \sigma_N^{(\alpha)}, \quad \hat \sigma_i^{(\alpha)} \in \{ \hat x, \hat y, \hat z, \hat 1 \},
\end{align}
as unitary generators. A feature which distinguishes the iQCC method from other adaptive approaches is the usage of an unrestricted operator pool. The iQCC operator pool is taken to be the set of all $4^N - 1$ possible $N$ qubit Pauli products, where structure in the qubit-mapped electronic Hamiltonian $\hat H$ and reference state facilitate an operator ranking procedure which scales polynomially in $N$. \cite{iQCC} Following optimization of the generators' associated variational amplitudes, the unitary is used to transform the current step Hamiltonian in the qubit space. The iQCC method has demonstrated systematic convergence towards ground state and excited state energies for problems requiring up to $72$ logical qubits. \cite{iQCC, iQCC_complex}  Furthermore, \textit{a posteriori} corrections have been developed to further increase accuracy of iQCC energy estimates. \cite{RyabinkinENPT} The employed Hamiltonian transformations are exact, where formally all BCH terms are summed in acquiring $\hat H_{\rm{eff}}$ with no neglecting of high body fermionic terms. Hence, while iQCC is a formally variational unitary downfolding procedure, the effective iQCC Hamiltonians have demonstrated rapid increase in the number of terms, particularly accentuated for the initial iterations. It has been elucidated that the rate of Pauli product increase diminishes at later iterations. \cite{iQCC_complex} Nevertheless, the main computational expense of the iQCC procedure arises from the introduction of a prohibitive number of Pauli products in the iQCC effective Hamiltonians.

In this work, we consider screening the Pauli product operator pool for elements which not only guarantee energy lowering via their energy gradient importance measure, but elements which minimize onset of new Pauli products in the exactly transformed current step effective Hamiltonian. This is accomplished by introducing a novel operator pool sampling scheme for the iQCC methodology which is sensitive to the introduction of new Pauli products in the transformed Hamiltonian. Further, we introduce an efficient approach to identifying elements from the exponentially sized unrestricted operator pool which minimize the growth of the transformed Hamiltonian while possessing a fixed energetic gradient magnitude. To balance consideration of energetic gradients and the expected increase of terms when selecting from the operator pool, we introduce a parameterized scoring function assigned to pool elements which acts as a cumulative importance measure. 

The rest of the paper is structured as follows. In Section \ref{sec:theory}, we review the necessary aspects of the iQCC methodology, and introduce an operator pool importance measure which takes into account the expected growth of candidates for selection. In Section \ref{sec:assessment}, we perform numerical tests for the H$_4$, N$_2$, and H$_2$O systems in strongly correlated regimes. We provide a discussion and an outlook on future perspectives in Section \ref{sec:discussion}. 

\section{Theory} \label{sec:theory}

\subsection{Iterative Qubit Coupled Cluster}
The iterative qubit coupled cluster (iQCC) method is an adaptive variational quantum algorithm utilizing effective Hamiltonians which has demonstrated systematic convergence towards ground state energies. \cite{iQCC, iQCC_complex} The iQCC method involves iteratively unitarily transforming the qubit-space Hamiltonian $\hat H$ with parameterized unitary transformations.  \cite{iQCC} An iteration begins with selection of $N_{\rm{gen}}$ Pauli products $\{ \hat P_\alpha \}_{\alpha=1}^{N_{\rm{gen}}}$, which define variational ansatz 
\begin{align} \label{eq:qcc_ansatz}
\hat U_{\rm{QCC}} = \prod_{k=1}^{N_{\rm{gen}}} e^{- i \tau_\alpha \hat P_\alpha/2}. 
\end{align}
The energy is then optimized with respect to amplitudes $\boldsymbol{\tau} = \{ \tau_\alpha \}_{\alpha=1}^{N_{\rm{gen}}}$ as 
\begin{align} \label{eq:qcc_opt}
E_K = \min_{\boldsymbol{\tau}} \bra{0} \hat U_{\rm{QCC}}^\dagger(\boldsymbol{\tau}) \hat H^{(K)} \hat U_{\rm{QCC}} (\boldsymbol{\tau}) \ket{0},
\end{align}
where $\ket{0}$ is a reference computational basis state (a Slater determinant in the fermion-to-qubit mapped orbital basis), and $K=0,1 \hdots$ denotes the iteration number. Pauli products are included in $\hat U_{\rm{QCC}}$ based on the magnitude of the first derivative of the energy with respect to their associated variational amplitude,
\begin{align} \label{eq:gradmag}
g_\alpha & = \left| \frac{\partial E_K}{\partial \tau_\alpha} \Big|_{\tau_\alpha=0}  \right|,
\end{align}
where
\begin{align} 
\frac{\partial E_K}{\partial \tau_\alpha} \Big|_{\tau_\alpha=0} & = \frac{\partial}{\partial \tau_\alpha }\bra{0} e^{i \tau_\alpha \hat P_\alpha / 2} \hat H^{(K)} e^{-i \tau_\alpha \hat P_\alpha / 2} \ket{0}  \\
& =  - \frac{i}{2} \bra{0} [\hat H^{(K)}, \hat P_\alpha] \ket{0}.
\end{align}
Pauli products with the highest values of $g_\alpha$ are selected for inclusion in $\hat U_{\rm{QCC}}$. Further, the exponentials in Eq.(\ref{eq:qcc_ansatz}) are ordered by ascending gradient magnitude (the highest gradient term acts directly on the reference wavefunction).  Unlike  other adaptive VQE schemes featuring a predefined and restricted operator pool (such as generators arising from fermionic single and double excitations with respect to $\ket{0}$), the iQCC method utilizes the unrestricted pool of all non-trivial $N$-qubit Pauli products, $\{ \hat P_\alpha \}_{\alpha=1}^{4^N-1}$. Thus, to achieve an efficient screening of the operator pool, one can not directly evaluate $g_\alpha$ for all pool elements individually, as is typically done in other adaptive VQE schemes. Instead, utilizing structure in the qubit-mapped $\hat H$ and reference state $\ket{0}$, one can constructively obtain the elements in this unrestricted Pauli product pool with non-zero gradient $g_\alpha$ using a classical algorithm efficient in the number of qubits $N$ and number of Pauli products in the qubit-mapped Hamiltonian, $M$. We review the derivation of this procedure in Appendix \ref{app:qcc_grads}. To summarize the main result, the qubit-space Hamiltonian may be expressed as 
\begin{align} \label{eq:ising_H_maintext}
    \hat H = \sum_i \left( \sum_j c_{j}^{(i)} \hat Z_j^{(i)} \right) \hat X_i,
\end{align}
where $\hat Z_j^{(i)}$ ($\hat X_i$) are tensor products of Pauli $\hat z$ ($\hat x$) operators. We partition $\{\hat P_\alpha \}_{\alpha=1}^{4^N-1}$ under the equivalence relation: if Pauli products $\hat P_\alpha$ and $\hat P_\beta$ can be expressed (up to a phase) as  $\hat Z_\alpha \hat X_\alpha$ and $\hat Z_\beta \hat X_\alpha$ respectively, and that $\hat P_\alpha$ and $\hat P_\beta$ possess the same number of Pauli $\hat y$ operators modulo $2$, then $\hat P_\alpha \sim \hat P_\beta$. If $\hat P_\alpha \sim \hat P_\beta$, then $g_\alpha = g_\beta$. The gradient magnitude of Pauli product $\hat P_\alpha = \hat Z_\alpha \hat X_\alpha$ (ignoring phase) is zero unless two criteria are met: 1) the Pauli product possesses an odd number of Pauli $\hat y$ operations (the size of overlap of supports of $\hat Z_\alpha$ and $\hat X_\alpha$ is odd), and 2) $\hat X_\alpha = \hat X_i$ for some $\hat X_i$ in Eq. (\ref{eq:ising_H_maintext}). This lets us constructively characterize the non-zero gradient partitions of the Pauli product pool by classically computing the gradient magnitude for each $\hat X_i$ in Eq. (\ref{eq:ising_H_maintext}). Elements from the highest gradient partitions are selected for inclusion in $\hat U_{\rm{QCC}}$. We denote the set of unique non-zero gradient magnitudes as $\{ g^{(k)} \}_k$, where all $\hat P_\alpha$ in partition $k$ have $g_\alpha = g^{(k)}$.

Following optimization of the $\hat U_{\rm{QCC}}$ amplitudes, the next-step Hamiltonian is obtained as $\hat H^{(K+1)} = \hat U_{\rm{QCC}}^\dagger(\boldsymbol{\tau}') \hat H^{(K)} \hat U_{\rm{QCC}} (\boldsymbol{\tau}')$ using the optimal amplitudes, $\boldsymbol{\tau}' = \argmin_{\boldsymbol{\tau}} E_K$. This transformation is classically performed via recursion, where transformation generated by a single Pauli product gives
\begin{align} \label{eq:dressing}
e^{i \tau'_\alpha \hat P_\alpha/2} \hat H e^{- i \tau'_\alpha \hat P_\alpha/2} = & \hat H - \frac{i}{2} \sin{\tau'_\alpha}[\hat H, \hat P_\alpha]   \nonumber \\ 
& + \frac{1}{2} \left( 1 - \cos \tau'_\alpha \right) \left( \hat P_\alpha \hat H \hat P_\alpha - \hat H \right).
\end{align}
The next iQCC iteration begins with partition screening of the Pauli product pool with respect to $\hat H^{(K+1)}$, and the procedure is repeated until a convergence criterion is met, such as in $E_K$'s or via norm of $\{ g^{(k)} \}$.
 
\subsection{Assessing Hamiltonian growth} \label{subsec:growth}

For adaptive VQE schemes employing effective Hamiltonians, the number of Pauli products in the transformed Hamiltonian is a crucial performance metric, as classical and quantum measurement resource requirements will generally increase with $M$. However, the exact number of measurements required to estimate the energy to a desired precision will be dependent on the energy estimator being employed, as well as the current step trial wavefunction. Let $\mathcal{H}^{(K)} = \{ \hat P_i \}_{i=1}^M$ denote the set of Pauli products in $\hat H^{(K)}$. By inspection of Eq. (\ref{eq:dressing}), new terms in the Hamiltonian transformed with $\exp(- i \tau_\alpha^{'} \hat P_\alpha / 2)$ are introduced only by $[\hat H, \hat P_\alpha]$. Since two Pauli products either commute or anticommute, $\hat P_\alpha \hat H \hat P_\alpha$ will not introduce Pauli products not already in $\hat H$, since $\hat P_\alpha \sum_i c_i \hat P_i \hat P_\alpha = \sum_i c_i \theta_i \hat P_i$, where $\theta_i$ equals $1$ ($-1$)  if $\hat P_i$ and $\hat P_\alpha$ commute (anticommute). Further, any commutator $[\hat P_i, \hat P_\alpha]$ in $[\hat H, \hat P_\alpha] = \sum c_i [\hat P_i, \hat P_\alpha]$ will contribute a new term not found in $\mathcal{H}^{(K)}$ if $\hat P_i$ and $\hat P_\alpha$ anticommute and $[\hat P_i, \hat P_\alpha] = 2 \hat P_i \hat P_\alpha$ is not proportional to a term in $\mathcal{H}^{(K)}$. Therefore, we can quantify the growth incurred by transforming the Hamiltonian with $\exp(- i \tau_\alpha^{'} \hat P_\alpha / 2)$ as the number of terms in $[\hat H^{(K)}, \hat P_\alpha]$ that are not found in $\mathcal{H}^{(K)}$. We denote this quantity as $\gamma_\alpha$, which serves as an analytically exact upper bound to the increase in number of terms upon conjugating the current step Hamiltonian with $\exp(-i \tau_\alpha^{'} \hat P_\alpha / 2)$,
\begin{align} 
| \mathcal{H}^{(K+1)} | \leq | \mathcal{H}^{(K)}| + \gamma_\alpha.
\end{align}
The case of inequality arises when there are term cancellations, or when magnitude of a Pauli product coefficient in the resulting Hamiltonian falls below a precision threshold, and is hence omitted. Such cases will be dependent on the value of $\tau_\alpha^{'}$, which are not accounted for in the analytical growth consideration.

In the context of an adaptive Hamiltonian transformation, we generally wish to select Pauli products from the pool with high $g_\alpha$ and low $\gamma_\alpha$. Similarly to how energetic gradients can not be directly assessed for the exponential number of elements in the Pauli group, assessing growth explicitly for all pool elements is computationally expensive. Instead, we devise efficient classical algorithms for obtaining the pool element with smallest $\gamma_\alpha$ within a given gradient partition. These algorithms and their derivation are included in Appendix \ref{app:finding_normalizers}. For a given partition $k$, we denote the minimal growth by $\gamma^{(k)}$, i.e., the lowest growth observed in the set of Pauli products with gradient $g^{(k)}$. We can hence use both $g^{(k)}$ and $\gamma^{(k)}$ in determining which partition of the Pauli product pool will be sampled from. In the next section, we introduce a parameterized scoring function which ranks the partitions of the Pauli product pool based on both $g^{(k)}$'s and $\gamma^{(k)}$'s. 

\subsection{Hybrid operator importance measures}

The majority of adaptive ansatz schemes, including iQCC, have used proxies for expected energy lowering such as energetic gradients $g_\alpha$'s as the sole importance measure when considering which elements from the operator pool should be included in the ansatz at a given iteration. In the context of an effective Hamiltonian theory, we can introduce another importance measure based on the worst-case increase in the number of Hamiltonian Pauli products upon transformation generated by a given element. To consider both energetic gradients and growth in selecting from the operator pool, we introduce score function $s^{(k)}$ assigned to partitions of the pool. Since energetic convergence will largely be determined by the energetic gradients of the chosen generators, we only consider scoring the $P$ highest gradient pool partitions. We define this cumulative score function as
\begin{align} \label{eq:scoring_func}
    s^{(k)} = a \tilde{g}^{(k)} - \left( 1-a \right ) \tilde{\gamma}^{(k)},
\end{align}
where $\tilde{g}_k$ and $\tilde{\gamma}^{(k)}$ are normalized energetic gradients and growth, respectively,
\begin{align}
    \tilde{g}^{(k)} & = g^{(k)}  \left( \frac{\sum_{i=1}^P g^{(i)}}{P} \right)^{-1}, \\
    \tilde{\gamma}^{(k)} & = \gamma^{(k)}  \left( \frac{\sum_{i=1}^P \gamma^{(i)}}{P} \right)^{-1}.
\end{align}
Real parameter $a \in [0,1]$ in Eq. (\ref{eq:scoring_func}) acts as a bias for energetic gradient. For instance, sampling from the highest energy gradient partition regardless of minimum growth estimates is accomplished by setting $a=1$.

\section{Numerical assessments} \label{sec:assessment}
Within this section we benchmark utilization of iQCC effective Hamiltonians obtained with operator selection employing the hybrid importance measure Eq. (\ref{eq:scoring_func}) for polyatomic systems exhibiting strong correlation effects. We compare to the canonical sampling choice used in the conventional iQCC methodology. \cite{iQCC} To recall, a gradient partition of the Pauli group is the set of $2^{N-1}$ Pauli products possessing an odd number of $\hat y$ operations with identical $\hat X_\alpha$ in factorization $\hat P_\alpha = \hat Z_\alpha \hat X_\alpha$ omitting multiplicative phase. The canonical sample from partition characterized by $\hat X_\alpha$ is defined by placing a single $\hat y$ operation on the index of the first non-trivial qubit acted on by $\hat X_\alpha$, the remaining non-trivial qubit indices are populated with $\hat x$ operators. For example, $\hat y_1 \hat x_2 \hat x_3 \hat x_4$ is the canonical iQCC operator choice from a partition characterized by $\hat X_\alpha = \hat x_1 \hat x_2 \hat x_3 \hat x_4$. Below, we outline the procedure for \textit{growth mitigated} iQCC, denoted GM($a$)-iQCC, where $a$ denotes the gradient bias utilized in scoring function $s(a)$ (see Eq. (\ref{eq:scoring_func})). We use $s(a)$ with several different gradient bias values for assessment. All iQCC calculations herein use $N_{\rm{gen}} = 1$, hence the highest gradient partition is always selected from in the canonical scheme. Explicitly, the employed procedure for obtaining the $K^{\rm{th}}$ step effective Hamiltonian in the GM($a$)-iQCC scheme is:
\begin{enumerate}
    \item Obtain description of the Pauli group partitioning for $\hat H^{(K-1)}$ (see Appendix \ref{app:qcc_grads}). Let $\{ g^{(k)} \}_{k=1}^P$ be the set of the $P$ highest energetic gradient magnitudes Eq. (\ref{eq:gradmag}).
    \item For each of the $P$ highest gradient partitions, obtain the lowest growth incurring element (see Appendix \ref{app:finding_normalizers}). In doing so, we obtain the minimal growth value of the top $P$ partitions $\{ \gamma^{(k)} \}_{k=1}^P$.
    \item Compute the scores $\{ s^{(k)} \}_{k=1}^{P}$ of the top $P$ partitions using chosen bias $a$, and computed $\{ g^{(k)} \}_{k=1}^P$ and $\{ \gamma^{(k)} \}_{k=1}^P$ (see Eq. (\ref{eq:scoring_func})).
    \item Select $\hat P_K$ corresponding to the growth minimizing element of the top scoring partition according to $\{ s^{(k)} \}_{k=1}^P$. \label{item:select_growth_min_element}
    \item Perform optimization 
    \begin{align} 
    \min_{\tau_K} E_K(\tau_K) = \bra{0} e^{i \tau_K \hat P_K / 2} \hat H^{(K-1)} e^{-i \tau_K \hat P_K / 2} \ket{0}. 
    \end{align}
    Converged $\min_{\tau_K} E_K(\tau_K)$ is taken as the $K^{\rm{th}}$ step of the energy estimation.
    \item Using $\tau_K^{'} = \argmin E_K(\tau_K)$, obtain $K^{\rm{th}}$ step effective Hamiltonian $\hat H^{(K)}$ after transformation Eq. (\ref{eq:dressing}). 
\end{enumerate}

It should be emphasized that Step \ref{item:select_growth_min_element} features the selection of the \textit{growth minimizing element} of the highest scoring partition. For instance, the $s(1)$ parameterization results in the growth minimizing element of the highest energy gradient partition being selected for at each step in the GM($1$)-iQCC procedure. This differentiates GM($1$)-iQCC from iQCC utilizing the canonical sampling scheme, while also always selects from the highest gradient partition, however with no effort made to minimize growth by the specific Pauli product chosen within the partition. 

For all systems studied herein, the Jordan-Wigner encoding was utilized in obtaining fermion-to-qubit mapped Hamiltonians { through the \texttt{OpenFermion} package,\cite{openfermion}} with no qubit tapering procedures employed. Hence a fermionic Hamiltonian defined over $N_{\rm{mo}}$ active molecular orbitals (MOs) results in a qubit-space Hamiltonian of $N = 2 N_{\rm{mo}}$ qubits. All fermionic Hamiltonians were resolved in the restricted Hartree-Fock (RHF) MO basis. All iQCC calculations utilize a fixed reference state taken to be the qubit-space RHF state, which corresponds to a computational basis state under the mapping of RHF spin orbitals to qubits. 

\begin{figure}
  \centering
  \includegraphics[width=0.4\textwidth]{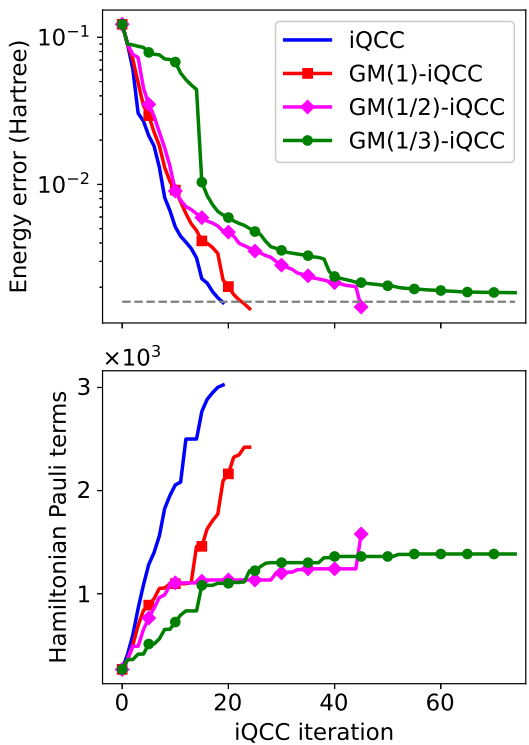}
  \caption{Comparison of iQCC procedure convergences applied to the H$_4$ chain in the STO-3G basis set with an equidistant H$-$H separation of $1.5 $ \AA{}. The energy error is the difference between the iQCC energy estimate at a given iteration and the FCI energy. The horizontal grey dashed line indicates an error of $1$ kcal/mol ($1.6$ milli-Hartree).}
  \label{fig:H4_convergences}
\end{figure}

\subsection{H$_4$ chain}
The electronic Hamiltonian for the linear H$_4$ chain was generated in the minimal STO-3G basis set with an equidistant H$-$H separation of $R=1.5 $ \AA. No freezing of orbitals was performed, resulting in a Hamiltonian defined over $4$ MOs and hence a $N=8$ qubit mapped Hamiltonian. Energy errors are taken with respect to the full configuration interaction (FCI) solution{, obtained via exact diagonalization}. In the GM($a$)-iQCC procedures, we use $r = \lceil \log_2 M \rceil$ in the growth minimizing search (see Algorithm \ref{alg:det_alg} of Appendix \ref{app:finding_normalizers}).

In Figure \ref{fig:H4_convergences}, convergence of iQCC energies utilizing various parameterizations of the partition score function are compared with the canonical sampling scheme. It is observed that operator selection via the $s(1)$ parameterization, utilized in GM($1$)-iQCC, gives a similar convergence trajectory to canonical iQCC selection, only requiring a few more iterations to achieve a desired accuracy of $1$ kcal/mol ($1.6$ milli-Hartree). This is expected, since in both schemes, the highest energy gradient partition is selected from for the current step Hamiltonian. However, while the schemes sample from the same partition for the initial Hamiltonian, this generally changes for later effective Hamiltonians which have different gradient values for the pool partitions. Hence we still expect slight deviations in their energy convergence trajectories. While selection using $s(1)$ demonstrates energetic convergence on par with that obtained using canonical selection, only a decrease of $\sim 20\%$ in the number of Pauli products in the effective Hamiltonians required for achieving an accuracy of $1$ kcal/mol is observed. However, comparing the procedures with fixed iteration number, there are several instances where sampling via $s(1)$ provides substantially reduced Pauli product counts for effective Hamiltonians, e.g., $K=10$, where the GM($1$)-iQCC effective Hamiltonian provides a $\sim 50\%$ reduction in the number of Pauli products in the effective Hamiltonian compared to the canonical selection scheme.

More significant reductions in effective Hamiltonian Pauli product counts are obtained using partition score function with non-zero weight placed on growth value. For instance, $s(1/2)$ provides an effective Hamiltonian with $\sim 48 \%$ less terms than canonical sampling to achieve an error of less than { $1$ kcal/mol}. However, since in general, the GM($1/2$)-iQCC procedure no longer samples from the highest gradient partition at each step, we observe that the rate of energetic convergence is decreased substantially. While offering the most reduced effective Hamiltonians at earlier iterations, the $s(1/3)$ selection fails to systematically converge the iQCC energy beyond an error of $1.84$ milli-Hartree. Further, the GM($1/3$)-iQCC energies are significantly higher than the energies obtained via the other selection schemes for the earlier iterations. We attribute the poor convergence of the $s(1/3)$ selection scheme to be a symptom of over-biasing the growth consideration. In this regime, generators are being selected more so by their expected onset of new Pauli products in the next-step effective Hamiltonian, rather than their energy gradients. Since energetic convergence will largely be dictated by the gradients of the selected generators, it is crucial to not under-weight the importance of energetic gradients $g^{(k)}$ in Eq. (\ref{eq:scoring_func}). 

\subsection{N$_2$ molecule}

We perform assessment of GM-iQCC applied to N$_2$ in an active space derived from the cc-pVDZ basis set. The active space is constituted by six electrons in six molecular orbitals ($6$e,$6$o), the minimal complete active space to describe the triple bond formation, resulting in a $N=12$ qubit-mapped Hamiltonian. The three highest energy occupied MOs and three lowest energy unoccupied MOs in the converged RHF solution form the active space. Energy errors are reported relative to the complete active space configuration interaction (CASCI) solution{, obtained via exact diagonalization of the active space Hamiltonian}. The GM($a$)-iQCC procedures utilize $r = \lceil \log_2 M \rceil$ in the growth minimizing search.

\begin{figure}
  \includegraphics[width=0.4\textwidth]{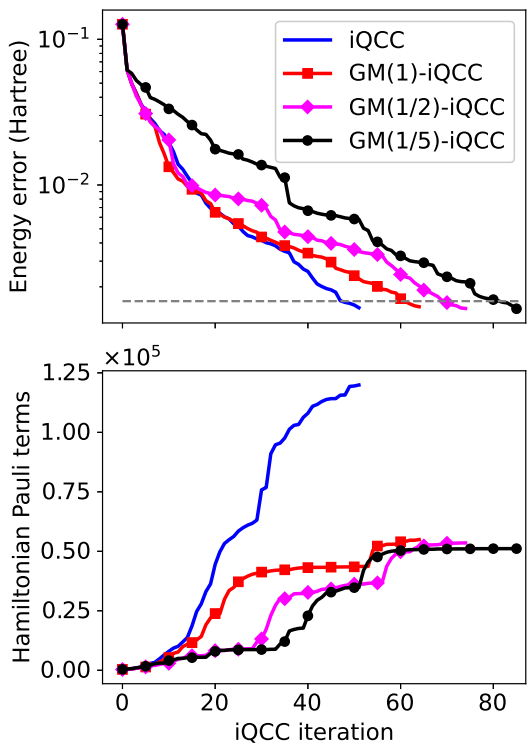}
  \caption{Comparison of iQCC procedure convergences applied to a CAS($6$e, $6$o) N$_2$ model in the cc-pVDZ basis set  with a bond distance of $1.5 $ \AA{}. The energy error is the difference between the iQCC energy estimate at a given iteration and the CASCI energy. The horizontal grey dashed line indicates an error of $1$ kcal/mol ($1.6$ milli-Hartree).}
  \label{fig:N2_convergence}
\end{figure}

In Figure \ref{fig:N2_convergence} we benchmark energetic convergences of iQCC procedures employing parameterized partition score function for selection along with the canonical selection at the stretched bond distance $R=1.5$ \AA{}. The canonical, $s(1)$, and $s(1/2)$ selection schemes give similar energetic convergence for the first $\sim 15$ iterations. As expected, iQCC and GM($1$)-iQCC share similar convergence trajectories, requiring $52$ and $65$ iterations to achieve { an error within $1$ kcal/mol}, respectively. Significantly, the effective $K=65$ GM($1$)-iQCC effective Hamiltonian consists of $5.5 \times 10^4$ terms, a $\sim 46 \%$ reduction compared to the $K=52$ effective iQCC Hamiltonian of $1.2 \times 10^5$ terms obtained via canonical selection. 

The $s(1/2)$ selection procedure provides drastically reduced term counts for effective Hamiltonians up to the $30^{\rm{th}}$ iteration, at which point a rapid onset of terms in the subsequent iterations brings the Pauli product counts to be comparable with the $s(1)$ procedure at similar iterations. By this iteration, all schemes' effective Hamiltonians provide substantially improved reference energies, however the GM($1$)-iQCC and GM($1/2$)-iQCC procedures offering substantially reduced effective Hamiltonian Pauli product counts. For instance, at $K=20$, the canonical selection provides an effective Hamiltonian of $3.6 \times 10^4 $ Pauli products and an energy error of $6.9$ milli-Hartree, while the $s(1/2)$ selection gives an effective Hamiltonian of $7.6 \times 10^3$, demonstrating a $\sim 79 \%$ reduction, with a comparable energy error of $8.7$ milli-Hartree.

\begin{figure*}
  \centering
  \includegraphics[width=0.95\textwidth]{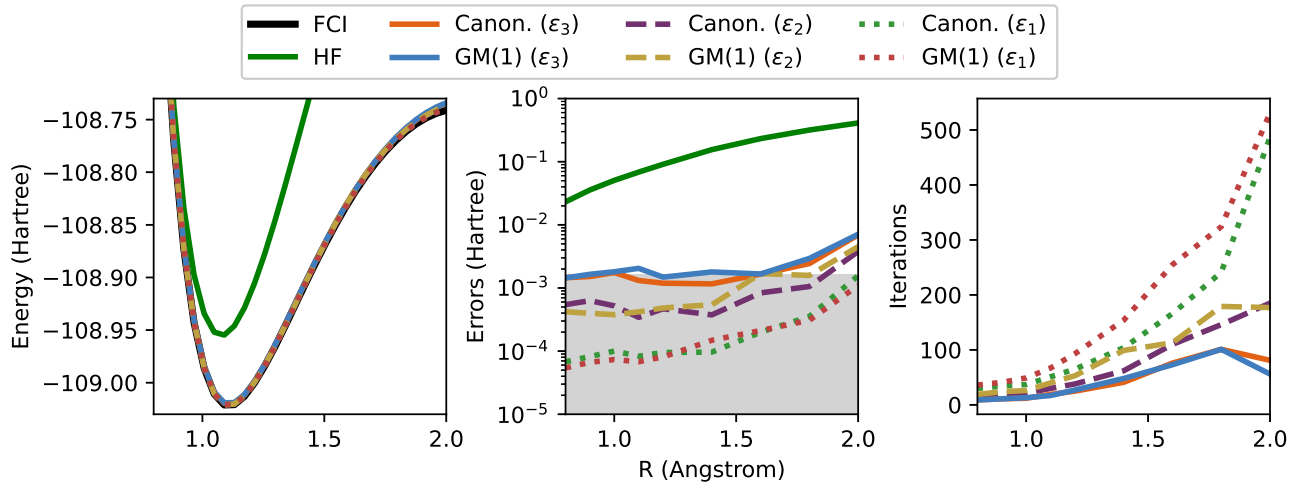}
  \caption{Potential energy surfaces obtained by the standard iQCC (``Canon.") and GM($1$)-iQCC procedures applied to the CAS($6$e, $6$o) N$_2$ model in the cc-pVDZ basis set. The energy error is the difference between the iQCC energy estimate at a given iteration and the CASCI energy. The grey shaded area of the middle plot indicates errors within $1$ kcal/mol ($1.6$ milli-Hartree) of the exact CASCI energy.}
  \label{fig:N2_pes}
\end{figure*}

While adaptive ansatz schemes inherently yield non-differentiable energy surfaces, it is desirable that the growth-mitigated iQCC relative errors over the considered bond lengths are not systematically worse than those of the standard iQCC procedure. To this end, potential energy surfaces PESs for the N$_2$ model obtained via iQCC and GM($1$)-iQCC procedures are shown in Figure \ref{fig:N2_pes}.  To assess the quality of PESs obtained via GM($1$) and standard iQCC procedures at various levels of convergence, we utilize a convergence criterion based on the gradient norm, i.e., the procedure exits when $\sum_k | g^{(k)} | \leq \epsilon_i$ for predefined threshold $\epsilon_i$. Figure \ref{fig:N2_pes} includes comparison of the GM($1$) and standard iQCC procedures at three gradient convergence thresholds: $\epsilon_1 = 0.1$, $\epsilon_2 = 0.2$, and $\epsilon_3 = 0.3$ Hartree.

For fixed convergence threshold, the standard and GM($1$) iQCC procedures demonstrate similar errors and number of required iterations at the majority of bond lengths. Utilizing thresholds $\epsilon_1$, $\epsilon_2$, and $\epsilon_3$, the standard iQCC procedures respectively achieve non-parallelity errors (NPEs) of $1.5 \times 10^{-3}$, $3.4 \times 10^{-3}$, and $5.7 \times 10^{-3}$, while the GM($1$) procedures obtain NPEs of $1.0 \times 10^{-3}$, $4.1 \times 10^{-3}$, and $5.6 \times 10^{-3}$. The NPE is defined as the difference between maximum error and minimum error relative to the exact CASCI energy obtained over all considered bond lengths. With considerably similar performance in terms of energetic errors and number of iterations required for convergence, it is to be noted that the advantage of the GM($1$) procedure lies in the reduced number of Hamiltonian terms along the iQCC procedure.

In Table \ref{table:n2_post_vqe}, benchmarking of simulated VQE application of iQCC and GM($a$)-iQCC pre-processed Hamiltonians is demonstrated, with comparisons to other classical electronic structure and VQE computations employing the original Hamiltonian. { All classical electronic structure calculations were carried out using the \texttt{Psi4} quantum chemistry package.\cite{Psi4}} The strongly correlated nature of the N$_2$ model with a bond length of $1.5$ \AA{} leads to a challenging case for the post-HF classical electronic structure methods, evident by the energetic errors of the coupled-cluster with singles and doubles (CCSD) and M\o ller-Plesset second order perturbation theory (MP2) energies being several milli-Hartree. Similarly, the dissociation of the linear H$_4$ chain to separated hydrogen atoms is a prototypical model exhibiting strong correlation effects. For the single point energy evaluation used for assessment, we set the equidistant H$-$H separation as $1.5$ \AA{} and utilize the STO-3G minimal basis. { Furthermore, benchmarking against the VQE employing the UCCSD ansatz as implemented in the \texttt{Tequila} package is performed.} The UCCSD ansatz likewise fails to achieve accuracy within { $1$ kcal/mol} for these problem instances. All methods involving VQE computations utilizing iQCC or GM($a$)-iQCC preprocessed qubit space Hamiltonian feature transformation including only a single generator per iteration. For example, method ``GM($1$)-iQCC($10$)'' in Table \ref{table:n2_post_vqe} denotes a preprocessed Hamiltonian obtained after twenty iterations of the iQCC procedure employing the $s(1)$ scoring function for the single generator selection performed at each iteration. Subsequently, the QCC method utilizing ten generators is applied to the final effective Hamiltonian. The energy error reported is the difference between the final optimized QCC energy and the CASCI energy. In general, the iQCC preprocessed Hamiltonians allow for much shallower quantum circuits, at the expense of more Pauli products in the Hamiltonian entering the QCC procedure. The GM($a$)-iQCC procedure likewise produces quantum resource reductions competitive with iQCC, while offering substantially reduced effective Hamiltonian Pauli product counts.

\begin{table*}
\caption{Comparison of methods applied to models in the strongly correlated regime, namely CAS($6$e, $6$o) N$_2$ in the cc-pVDZ basis set and a linear H$_4$ chain in the STO-$3$G basis set. Both the N$-$N bond distance and the equidistant H$-$H separation were set to $R=1.5$ \AA{}, where both systems possess a high degree of orbital degeneracies. The HF, MP2, and CCSD computations were carried out using the \texttt{Psi4} package. \cite{Psi4}  The UCCSD  calculation were performed using the \texttt{Tequila} package, where the cluster operator was obtained in closed-shell form before a first-order Trotterization. \cite{Tequila} QCC($N_\text{ent}$) denotes the QCC ansatz constructed from $N_{\text{ent}}$ selected generators. (GM($a$)-)iQCC($N_{\text{ent}}$) denotes the QCC($N_{\text{ent}}$) procedure using \mbox{(GM($a$)-)iQCC} preprocessed Hamiltonian. All preprocessed Hamiltonians were obtained after $K=20$ iterations. Errors are reported in milli-Hartree. The number of Pauli products in the Hamiltonian utilized in state vector VQE simulation is denoted by $M$. The reported CNOT counts were obtained via circuit compilation using the \texttt{Tequila} package with the \texttt{Qiskit} backend. \cite{Qiskit}} 
\centering
\footnotesize
\def\sym#1{\ifmmode^{#1}\else\(^{#1}\)\fi}%
\begin{tabular}{l*{8}{c}}
\toprule
\rule{0pt}{2.5ex}Method & \multicolumn{4}{c}{N$_2$} & \multicolumn{4}{c}{H$_4$} \\ \cmidrule(lr){2-5} \cmidrule(lr){6-9}
& Error & $M$ & CNOTs & Circuit parameters & Error & $M$ & CNOTs & Circuit parameters \\ \hline 
HF & $192$ & - & - & - & $167$ & - & - & -\\ 
CCSD & $8.02$   & - & - & - & $-1.47$ & - & - & -\\
MP2  & $-16.8$ & - & - & - & $80.5$ & - & - & - \\
UCCSD                      & $2.84$   & $247$                   & $4512$                  & $30$    & $2.59$ & $185$ & $1280$ & $12$                \\
QCC($10$)                     & $14.6$  & $247$                  & $60$                    & $10$    & $3.30$ & $185$ & $60$  & $10$                \\
iQCC($10$)     & $4.66$  & $36281$                 & $88$                    & $10$    &   $0.522$ & $3024$ & $80$ & $10$              \\
\rule{0pt}{2.5ex}GM($1$)-iQCC($10$)    & $4.62$  & $21992$                 & $68$                    & $10$    &  $0.825$ & $2096$ & $84$ & $10$              \\
\rule{0pt}{2.5ex}GM($\frac{1}{2}$)-iQCC($10$) & $7.48$  & $8567$                  & $92$                    & $10$       & $1.62$ & $1133$ & $72$ & $10$             \\

QCC($50$)                     & $1.69$  & $247$                   & $300$                   & $50$ & $0.161$ & $185$ &  $252$ & $50$               \\
iQCC($50$)     & $0.670$ & $36281$                 & $452$                   & $50$  & $4.36 \times 10^{-2}$ &      $3024$ &  $300$ & $50$            \\
\rule{0pt}{2.5ex}GM($1$)-iQCC($50$)     & $0.814$ & $21992$                 & $416$                   & $50$  & $5.00 \times 10^{-3}$ & $2096$ & $328$ & $50$                \\
\rule{0pt}{2.5ex}GM($\frac{1}{2}$)-iQCC($50$) & $1.22$  & $8567$                  & $436$                   & $50$    & $4.65 \times 10^{-3}$ & $1133$ & $316$ &    $50$             \\
\hline
\end{tabular}
\label{table:n2_post_vqe}
\end{table*}
 
\subsection{H$_2$O molecule}
As a larger scale example, we compare performance of the first few iterations of iQCC and GM($1$)-iQCC for H$_2$O in the 6$-$31G(d) basis set. The lowest-energy MO, corresponding to the $1s$ atomic orbital of the oxygen atom, was frozen, with the remaining $18$ MOs constituting the active space, yielding a $36$ qubit Hamiltonian. A symmetric O$-$H bond distance of $1.5 $ \AA{} was used, with H$-$O$-$H angle set to $107.60^{\circ}$. The initial qubit space Hamiltonian, obtained via the Jordan-Wigner transformation of the fermionic Hamiltonian in the canonical orbital basis, contains $41915$ Pauli products. To assess the relative quality of iQCC energies obtained via the canonical and hybrid operator selection scheme, we obtain energetic errors with respect to the CASCI energy, obtained via Davidson diagonalization in the determinantal basis using the \texttt{GAMESS} package. \cite{gamess}

We compare the canonical iQCC and GM($1$)-iQCC procedures in Table \ref{tab:h2o_results}. In this instance, we use $r = \left \lceil M/10 \right \rceil$, as the growth minimizing search using $r = \lceil \log_2 M \rceil$ could not find the growth minimizing element in the highest gradient partition for the initial Hamiltonian. This can arise when there exists a Pauli product in the searched partition which incurs a low growth as a result from commuting with an anomalously large number of terms in the Hamiltonian. For the first six iterations, the canonical choice and the growth minimizing choice found within the highest gradient partition via the $s(1)$ procedure provide identical optimized trial energies up to $10^{-4}$ Hartree. The reduced proliferation of new Pauli products in the GM($1$)-iQCC procedure becomes pronounced as further iterations are performed. For the eighth iteration, a $43.5$ percent reduction in terms using the $s(1)$ selection procedure ($2.17 \times 10^{5}$ Pauli products) compared to the canonical selection ($3.84 \times 10^5 $ Pauli products) is observed. Furthermore, the generators selected along the $s(1)$ iQCC procedure also provide a slightly lower optimized energy by the eighth iteration compared to the iQCC procedure employing the canonical selection. 

\begin{table}[H]
    \caption{Comparison of the iQCC and GM($1$)-iQCC procedures applied to the $36$ qubit H$_2$O model in the 6$-$31G(d) basis with symmetric bond length of $1.5$ \AA{}. Numerics for the iQCC and GM($1$)-iQCC methods for a given iteration are tabulated in uncoloured and grey rows respectively.}
    \centering
    \begin{tabular}{|c|c|c|}
    \hline 
         Iteration & Energy error (Hartree) & Pauli products \\
    \hline
         \multirow{2}{*}{$1$} & $0.2319$ & $5.63 \times 10^4$  \\
         & \cellcolor{gray!25} $0.2319$ & \cellcolor{gray!25} $5.63 \times 10^4$ \\ \hline 
         \multirow{2}{*}{$2$} & $0.2293$ & $6.79 \times 10^4$ \\
         & \cellcolor{gray!25} $0.2293$ & \cellcolor{gray!25} $6.65 \times 10^4$ \\ \hline 
          \multirow{2}{*}{$4$} & $0.2092$ & $9.63 \times 10^4$ \\
         & \cellcolor{gray!25} $0.2092$ & \cellcolor{gray!25} $9.12 \times 10^4$ \\ \hline 
         \multirow{2}{*}{$6$} & $0.2048$ & $1.82 \times 10^5$ \\
         & \cellcolor{gray!25} $0.2048$ & \cellcolor{gray!25} $1.41 \times 10^5$ \\ \hline 
         \multirow{2}{*}{$8$} & $0.2023$ & $3.84 \times 10^5$ \\
         & \cellcolor{gray!25} $0.2016$ & \cellcolor{gray!25} $2.17 \times 10^5$ \\
    \hline
    \end{tabular}
    \label{tab:h2o_results}
\end{table}

\section{Discussion} \label{sec:discussion}

In this work, we have introduced an operator pool importance measure which takes into account the expected onset of terms in the iteratively transformed Hamiltonian. The hybrid importance measure, which has contributions from expected growth and energetic gradients, was utilized in the selection of iQCC generators for molecular systems in strongly correlated regimes. Due to the exponential size of the operator pool employed, i.e., the $N$-qubit  Pauli group, efficient algorithms to finding growth-minimizing elements were devised which avoid exhaustively computing the growth for all possible elements. The benefit of utilizing the full Pauli group as the operator pool is the potential for selecting Pauli products which occur in arbitrary rank fermionic excitations. Furthermore, it circumvents the problem of ``barren plateaus'' in energy optimization,\cite{McClean2018} since the selected generators are always of non-zero gradient by construction, and this selection is exhaustive over a complete operator basis. It was found that so long as biasing towards energetic gradients is not made too small in the pool partition score function, we retain robust energetic convergence towards FCI and CASCI energies, while achieving significant reductions in the number of resulting Hamiltonian Pauli products. Energetic convergences when using score function $s(1)$ were on par with the canonical iQCC selection procedure, while the former achieved up to a $\sim 50\%$ reduction in the number of resulting Hamiltonian Pauli products to achieve an energy error within $1$ kcal/mol. Placing greater bias towards minimizing growth can substantially reduce the size of early iteration effective Hamiltonians, however such a bias can impede energy lowering at later iterations. We hence consider the GM($1$)-iQCC procedure to be the most competitive approach to mitigating Hamiltonian growth while substantially reducing quantum resource requirements. Similar to the canonical iQCC procedure, the Hermitian character and eigenspectrum are formally conserved in the effective Hamiltonians, differentiating this approach from the large majority of efforts in VQE Hamiltonian preprocessing. However, the GM($1$)-iQCC procedure offers substantially smaller effective Hamiltonians compared to iQCC, while yielding competitive energies for fixed iteration count.

\section{Acknowledgements}
The authors thank Ilya G. Ryabinkin for many stimulating discussions, and providing computational details for the H$_2$O model studied in Section \ref{sec:assessment}. A.F.I. acknowledges financial support from the Google Quantum Research Program, Early
Researcher Award, and Zapata Computing Inc. This research was enabled in part by support provided by Compute Ontario and Compute Canada. 

\begin{appendix}

\section{Evaluating gradients of the unrestricted Pauli pool} \label{app:qcc_grads}

In Ref. \citenum{iQCC}, it was shown that the gradient absolute magnitudes of the $O(4^{N})$ operators in the unrestricted Pauli product pool can be obtained using resources scaling polynomially in the number of qubits and number of Hamiltonian Pauli products. The gradient evaluations Eq. (\ref{eq:gradmag}) are performed purely classically, further differentiating this pool from others previously employed in adaptive VQE schemes. For completeness, we review herein a derivation of the procedure used in the QCC method to accomplish this task.

To recall, the qubit-mapped Hamiltonian may be expressed as in Eq.(\ref{eq:ising_H_maintext}), where $\hat Z_j^{(i)}$ and $\hat X_i$ are Pauli products consisting of strictly Pauli $z$ and Pauli $x$ operators, respectively. This recasting of $\hat H$ is possible since an arbitrary Pauli product may be expressed as a product of Pauli $z$ and Pauli $x$ operators up to a sign. Note that, when $\hat H$ is purely real-valued, every $\hat Z_j^{(i)} \hat X_j$ product has an even number of Pauli $y$ operations and hence a real prefactor, $1$ or $-1$. The number of unique $\hat X_i$'s occurring in Eq. (\ref{eq:ising_H_maintext}) is proportional to the number of single and double excitations occurring in the initial Hamiltonian's fermionic form, i.e. scaling as $O(n^4)$. Inserting Eq. (\ref{eq:ising_H_maintext}) into the gradient expression for Pauli product $\hat P_\alpha$:
\begin{align} 
g_\alpha = \left| \Im \left( \sum_i \bra{0} \left( \sum_j c_{j}^{(i)} \hat Z_j^{(i)} \right) \hat X_i \hat P_\alpha \ket{0} \right) \right|,
\end{align}
where we have used the fact that $- \frac{i}{2} \bra{0} [\hat H, \hat P_\alpha] \ket{0} = \Im(\bra{0} \hat H \hat P_\alpha \ket{0})$ due to hermiticity of $\hat H$ and $\hat P$. The summation $\sum_j c_j^{(i)} \hat Z_j^{(i)}$ can be seen acting directly on the reference dual state $\bra{0}$, which is the adjoint of a mutual eigenstate of the strictly Pauli $z$ products, giving
\begin{align} 
g_\alpha = \left| \Im \left( \sum_i \left( \sum_j c_{j}^{(i)} 
\lambda_j^{(i)} \right) \bra{0}  \hat X_i \hat P_\alpha \ket{0} \right) \right|,
\end{align}
with $\hat Z_j^{(i)}\ket{0} = \lambda_j^{(i)} \ket{0}$ where $\lambda_j^{(i)} \in \{1, -1\}$. The imaginary component of expectation value $\bra{0} \hat X_i \hat P_\alpha \ket{0}$ must be vanishing unless $\hat P_\alpha$ meets certain requirements. Let
\begin{align} 
\hat X_i & = \prod_{p=1}^{n} \hat x_p^{\mu_p^{(i)}}, \\ 
\hat Z_i & = \prod_{p=1}^{n} \hat z_p^{\nu_p^{(i)}},
\end{align}
where $\mu_p^{(i)}, \nu_p^{(i)} \in \{ 0, 1 \}$. Expressing $\hat P_\alpha$ as $\theta_\alpha \hat X_\alpha \hat Z_\alpha$ with $\theta_\alpha = e^{i \pi \sum_{p=1}^n \mu_{p}^{(\alpha)} \nu_{p}^{(\alpha)} / 2 }$ ($\theta_\alpha$ is introduced to cancel any prefactor introduced in multiplication of factors $\hat X_\alpha$ and $\hat Z_\alpha$), 
\begin{align}
g_\alpha & = \left|\Im \left( \theta_\alpha \sum_i \left( \sum_j c_{j}^{(i)} 
\lambda_j^{(i)} \right)  \bra{0}  \prod_{p=1}^{n} \hat x_p^{\mu_p^{(i)}} \hat x_p^{\mu_p^{(\alpha)}} \hat Z_\alpha \ket{0} \right) \right|, \\
& = \left| \Im \left( \lambda_\alpha \theta_\alpha \sum_i \left( \sum_j c_{j}^{(i)} 
\lambda_j^{(i)} \right) \bra{0}  \prod_{p=1}^{n} \hat x_p^{\mu_p^{(i)}} \hat x_p^{\mu_p^{(\alpha)}} \ket{0} \right) \right|, \\
& = \left| \Im \left( \lambda_\alpha \theta_\alpha \sum_i \left( \sum_j c_{j}^{(i)} 
\lambda_j^{(i)} \right)  \prod_{p=1}^n \delta_{\mu_p^{(i)}, \mu_p^{(\alpha)}} \right) \right|,
\end{align}
where $\hat Z_\alpha \ket{0} = \lambda_\alpha \ket{0}$ and $\delta_{i,j}$ is the Kronecker delta function. Since $c_j^{(i)}$ are strictly real, the only quantity with potentially nonzero imaginary part is $\theta_\alpha$, hence
\begin{align}
g_\alpha & = \left| \Im (\theta_\alpha) \lambda_\alpha  \sum_i \left( \sum_j c_{j}^{(i)} 
\lambda_j^{(i)} \right)  \prod_{p=1}^n \delta_{\mu_p^{(i)}, \mu_p^{(\alpha)}} \right|, \\
& = \left| \Im(\theta_\alpha) \lambda_\alpha  \sum_i \left( \sum_j c_{j}^{(i)} 
\lambda_j^{(i)} \right)  \boldsymbol{\delta}_{\vec{\mu}^{(i)}, \vec{\mu}^{(\alpha)}} \right|,
\end{align}
where $\vec{\mu}^{(j)} \equiv (\mu_1^{(j)} ,\hdots \mu_n^{(j)})$, and $\boldsymbol{\delta}_{\vec{i}, \vec{j}}$ is the multidimensional Kronecker delta function, i.e. $\boldsymbol{\delta}_{\vec{i}, \vec{j}} = 1$ if $\vec{i}$ and $\vec{j}$ are identical vectors, otherwise $\boldsymbol{\delta}_{\vec{i}, \vec{j}} = 0$. 

Since $\Im(\theta_\alpha) = \sin(\pi \sum_{p=1}^n \mu_p^{(\alpha)} \nu_{p}^{(\alpha)} / 2) $, it is evident that $g_\alpha$ is vanishing unless  $\hat P_\alpha$ possesses an odd number of Pauli $\hat y$ operators. Furthermore, given $\hat P_\alpha$ with odd number of Pauli $\hat y$'s, $g_\alpha$ is uniquely determined by $\vec{\mu}^{(\alpha)}$:
\begin{enumerate} 
\item If $\vec{\mu}^{(\alpha)}$ is not identical to any $\vec{\mu}^{(i)}$ arising from the set of products of Pauli $x$ operators in Eq. (\ref{eq:ising_H_maintext}), $\{ \hat X_i \}_i$, i.e. $\boldsymbol{\delta}_{\vec{\mu}^{(i)}, \vec{\mu}^{(\alpha)}} = 0 \> \forall \> i$, then $g_\alpha = 0$. 
\item Otherwise, if $\boldsymbol{\delta}_{\vec{\mu}^{(i)}, \vec{\mu}^{(\alpha)}} = 1$ for some $i$, then $g_\alpha = \left|\sum_j c_j^{(i)} \lambda_j^{(i)} \right|$.
\end{enumerate}

Therefore, to efficiently check which elements of the unrestricted pool $\{ \hat P_\alpha \}_{k=1}^{4^n-1}$ have non-zero gradient: obtain the set of $\{ \vec{\mu}^{(i)} \}_i$, i.e. the binary vectors characterizing the unique Pauli $\hat x$ products arising in Eq. (\ref{eq:ising_H_maintext}). There are $O(n^4)$ unique $\hat X_i$ in Eq. (\ref{eq:ising_H_maintext}), hence $|\{ \vec{\mu}^{(i)} \}_i| = O(n^4)$ for the initial Hamiltonian. The corresponding absolute magnitude of the gradient for any Pauli product with  $\vec{\mu}^{(i)}$ in this set, and an odd number of $\hat y$ operations, is then computed as $ \left| \sum_j c_j^{(i)} \lambda_j^{(i)} \right|$ classically. Any choice of $\vec{\nu}^{(\alpha)}$, i.e. $\hat Z_\alpha$ in factorized $\hat P_\alpha = \hat Z_\alpha \hat X_\alpha$, will only potentially induce a sign change of $\lambda_\alpha$, and hence the absolute magnitude is invariant to the choice of $\vec{\nu}^{(\alpha)}$, so long as an odd number of Pauli $\hat y$ operators occur, i.e. $\vec{\mu}^{(i)} \cdot \vec{\nu}^{(\alpha)} \mod 2 = 1$.  

The set $\left \{ \left( \vec{\mu}^{(i)}, g_i \right) \right \}_i$ is a complete characterization of the non-zero gradient elements of $\{ \hat P_\alpha \}_{k=1}^{4^n-1}$. This set informs us that any Pauli product possessing an odd number of Pauli $\hat y$ operations with an $x$-string given by $\vec{\mu}^{(i)}$ must have gradient magnitude $g_i$. Therefore, to obtain a Pauli product $\hat P_\alpha$ with gradient $g_i$, define $\hat P_\alpha$ with $\vec{\mu}^{(i)}$ and choose any $\vec{\nu}^{(\alpha)}$ such that $\vec{\mu}^{(i)} \cdot \vec{\nu}^{(\alpha)} \mod 2 = 1$.

There are an exponential number of choices for satisfactory $\vec{\nu}^{(\alpha)}$, hence there is an exponential number of Pauli products for each unique gradient magnitude $g^{(i)}$. In previous QCC works, the ``canonical" choice of $\vec{\nu}^{(\alpha)}$ is defined by having a Pauli $\hat y$ operation on the qubit corresponding to the first non-zero component in $\vec{\mu}^{(\alpha)}$, with Pauli $\hat x$ operations placed at the remaining non-zero components. This choice was motivated by lowering circuit depths: it possesses the minimum number of non-trivial qubit operations required to have gradient magnitude $g^{(i)}$. However, this choice is not necessarily optimal in the context of Hamiltonian transformation. We describe an approach to finding growth-minimizing elements of fixed gradient magnitude in the following section.

\section{Obtaining growth-minimizing elements of fixed gradient from the Pauli group} \label{app:finding_normalizers}

This Appendix is organized as follows. Firstly, we formulate the conditions for finding Pauli products which do not introduce any new terms in $\hat H$ upon similarity transformation with $\exp(- i \tau_\alpha \hat P_\alpha / 2)$. We refer to such Pauli products as ``normalizer" Pauli products. Then, we propose a convenient method of computing the analytical growth of candidate Pauli products, which allows us to find normalizers if they exist for a particular Hamiltonian. If normalizers do not exist, we can use the methods developed herein to identify growth minimizing Pauli products. We then introduce deterministic and probabilistic algorithms for obtaining minimal growth Pauli products with fixed energetic gradient $g_\alpha$ values.

Since new terms arise strictly from commutator $[\hat H, \hat P_\alpha]$, the condition for $\hat P_\alpha$ to be a normalizer Pauli product is
\begin{align} \label{eq:normalizer_def}
    [\mathcal{H}, \hat P_\alpha] \subseteq \mathcal{H} \cup \{ 0 \},
\end{align}
where $\mathcal{H}$ is the set of Pauli products in $\hat H$, $\mathcal{H} = \{ \hat P_i \}_{i=1}^M$. The commutator of $\mathcal{H}$ and $\hat P_\alpha$ is defined to produce a set of element-wise commutators, i.e. $[\mathcal{H}, \hat P_\alpha] = \{ [\hat P_1, \hat P_\alpha], \hdots, [\hat P_M, \hat P_\alpha]  \}$. Here and throughout this section, we ignore coefficients arising in non-zero commutators between Pauli products. Let $\mathcal{H}_C^{(\alpha)} \subseteq \mathcal{H}$ denote the subset of Hamiltonian Pauli products which commute with $\hat P_\alpha$,
\begin{align}
    \mathcal{H}_C^{(\alpha)} = \{ \hat P_i \in \mathcal{H} \> | \> [\hat P_i, \hat P_\alpha] = 0 \},
\end{align}
and similarly let $\mathcal{H}_A^{(\alpha)} = \mathcal{H} / \mathcal{H}_C^{(\alpha)}$ denote the subset of Hamiltonian terms which anticommute with $\hat P_\alpha$. By definition, $[\mathcal{H}_C^{(\alpha)}, \hat P_\alpha] = \{ 0 \}$. Commutation of any $\hat P_i \in \mathcal{H}_A^{(\alpha)}$ with $\hat P_\alpha$ results in $[\hat P_i ,\hat P_\alpha] = 2\hat P_i \hat P_\alpha$, which anticommutes with $\hat P_\alpha$. Therefore condition Eq. (\ref{eq:normalizer_def}) can be refined as 
\begin{align}
[ \mathcal{H}_A^{(\alpha)}, \hat P_\alpha] = \mathcal{H}_A^{(\alpha)}. 
\end{align}
Essentially, normalizer $\hat P_\alpha$ induces an automorphism $[\cdot, \hat P_\alpha]$ on the set of Hamiltonian terms anticommuting with $\hat P_\alpha$. Since $[\hat P_i, \hat P_\alpha] = 2 \hat P_i \hat P_\alpha$ for $\hat P_i \in \mathcal{H}_A^{(\alpha)}$, it follows that there exists $\hat P_j \in \mathcal{H}_A^{(\alpha)}$ which is proportional to $\hat P_i \hat P_\alpha$, if $\hat P_\alpha$ is a normalizer Pauli product. Hence, for normalizer $\hat P_\alpha$, we must find $|\mathcal{H}_A^{(\alpha)} |/2$ disjoint pairs of terms $(\hat P_i, \hat P_j)$ such that $\hat P_i \hat P_j = \hat P_\alpha$ up to a multiplicative phase. Since $\hat P_j = \hat P_i \hat P_\alpha$ up to a phase, it also follows that $(\hat P_i, \hat P_j)$ are an anticommutative pair, and $\hat P_\alpha = \hat P_i \hat P_j$ anticommutes with both $\hat P_i$ and $\hat P_j$. These facts allow us to derive a classical algorithm to efficiently find growth minimizing terms, as described below.

The principal object used in formulating the algorithm is the multiset of non-zero commutators which can be evaluated between unique pairs formed from set $\mathcal{H}$. Let 
\begin{align} \label{eq:commutator_set}
    \mathcal{C} = \left\{ [\hat P_i, \hat P_j] \> | \> \hat P_i, \hat P_j \in \mathcal{H}, i < j \right\} / \{ 0 \},
\end{align}
where we again ignore any prefactors arising in nonzero $[\hat P_i, \hat P_j]$. The requirement of $i <j$ is used to ignore redundant pairs. Once $\mathcal{C}$ is evaluated, we can then check for the multiplicity of each unique $\hat P_\alpha \in \mathcal{C}$ (how many times $\hat P_\alpha$ occurs in $\mathcal{C}$), and denote this multiplicity as $m_{\mathcal{C}}(\hat P_\alpha)$. If one has $m_{\mathcal{C}}(\hat P_\alpha) = |\mathcal{H}_A^{(\alpha)}|/2$, then $\hat P_\alpha$ is a normalizer. Hence, with $m_{\mathcal{C}}(\hat P_\alpha)$ and $\mathcal{H}_A^{(\alpha)}$ for all unique $\hat P_\alpha \in \mathcal{C}$ computed, one can determine which $\hat P_\alpha \in \mathcal{C}$ are normalizers. Further, it can be shown that the multiplicity for $\hat P_\alpha$ is upper bounded by $m_{\mathcal{C}}(\hat P_\alpha) \leq |\mathcal{H}_A^{(\alpha)}|/2$, with equality met for normalizers. In any case, the number of new terms introduced to the Hamiltonian upon transformation generated by $\hat P_\alpha \in \mathcal{C}$ can be equated to 
\begin{align}
\gamma_\alpha = |\mathcal{H}_A^{(\alpha)}| - 2 m_{\mathcal{C}}(\hat P_\alpha),
\end{align}
where $\gamma_\alpha$ is the growth incurred by $\hat P_\alpha$. In essence, this is the general structure used in finding Pauli products with minimal growth. However, computing $|\mathcal{H}_A^{(\alpha)}|$ for all $\hat P_\alpha \in \mathcal{C}$ scales as $O(M^3)$, which can be prohibitively expensive for large problem instances. Further, we generally wish to find $\hat P_\alpha$ not only with low $\gamma_\alpha$ values, but also with high energy gradients $g_\alpha$. We present below approximate deterministic and probabilistic algorithms for obtaining low growth Pauli products with a fixed energy gradient. 

\subsubsection{Approximate deterministic algorithm}
We aim to find a Pauli product with the lowest $\gamma_\alpha$ that is in a given non-zero gradient partition. As reviewed in Appendix \ref{app:qcc_grads}, all Pauli products within a partition are characterized by an ``$x$-string" $\vec \mu^{(k)}$ with identical gradient magnitude $g^{(k)}$. We refer to the lowest $\gamma_\alpha$ found for this set as $\gamma^{(k)}$. The Pauli product corresponding to $\gamma^{(k)}$ is hence the Pauli product with gradient magnitude $g^{(k)}$ possessing the lowest growth. In essence, we formulate an efficient procedure for obtaining $\gamma^{(k)}$ and the corresponding Pauli product given $\vec \mu^{(k)}$. Since a Pauli product is uniquely defined by $\vec \mu$ and $\vec \nu$, this procedure can be seen as obtaining $\vec \nu$ which minimizes $\gamma$ for fixed $\vec \mu^{(k)}$. Herein we explain the algorithm in detail, and summarize the steps in Algorithm \ref{alg:det_alg}.

\vspace{1pt}
\begin{algorithm}[H]
  \caption{Deterministic search for low $\gamma_\alpha$ Pauli products with fixed gradient $g^{(k)}$. Let $\boldsymbol{\mu} \equiv \{ \vec{\mu}^{(i)} \}_i $ be the set of unique $x$-strings found in terms of $\mathcal{H}$. Let $\mathcal{Z}_i \equiv \{ \hat Z^{(i)}_l \}_l$ (see Eq. (\ref{eq:ising_H_maintext})). As in the description in the main text, non-unital coefficients on Pauli product commutator expressions are ignored.}
  \label{alg:det_alg}
  \hspace*{\algorithmicindent} \textbf{Input} $\mathcal{H}$, $\vec{\mu}^{(k)}$, $r$ \\
    \hspace*{\algorithmicindent} \textbf{Output} growth minimizing element $\hat P_\alpha$ with gradient $g^{(k)}$ and growth $\gamma^{(k)}$
   \begin{algorithmic}[1]
   \State{obtain set $\boldsymbol{\mu}$}
    \State{initiate $\mathcal{C} = \{ \}$}
   \For{$\vec{\mu}^{(i)}, \vec{\mu}^{(j)} \in \boldsymbol{\mu}$, $i < j$,}
        \If{$\vec{\mu}^{(i)} + \vec{\mu}^{(j)} \mod 2 = \vec{\mu}^{(k)}$}
            \For{$\hat Z_l^{(i)} \in \mathcal{Z}_i$, $\hat Z_{l'}^{(j)} \in \mathcal{Z}_j$}
                \State compute $\hat C = [\hat Z_l^{(i)} \hat X_i, \hat Z_{l^{'}}^{(j)} \hat X_j]$
                \If{$\hat C \neq 0$}
                    \State update $\mathcal{C} \to \mathcal{C} + \{ \hat C \}$
                \EndIf
            \EndFor
        \EndIf
   \EndFor
   \State count occurrences $m_{\mathcal{C}}(\hat P_\alpha)$ for each unique $\hat P_\alpha \in \mathcal{C}$
   \For{$r$ highest occurring $\hat P_\alpha \in \mathcal{C}$}
        \State compute $|\mathcal{H}_A^{(\alpha)}|$
        \State compute and save $\gamma_\alpha = |\mathcal{H}_A^{(\alpha)}| - 2 m_{\mathcal{C}}(\hat P_\alpha)$
   \EndFor 
   \State select $\hat P_{\alpha^{'}}$ with smallest $\gamma_{\alpha^{'}}$. Then $\gamma^{(k)} = \gamma_{\alpha^{'}}$
   \end{algorithmic}
\end{algorithm}

\begin{figure*}
  \centering
  \includegraphics[width=1\textwidth]{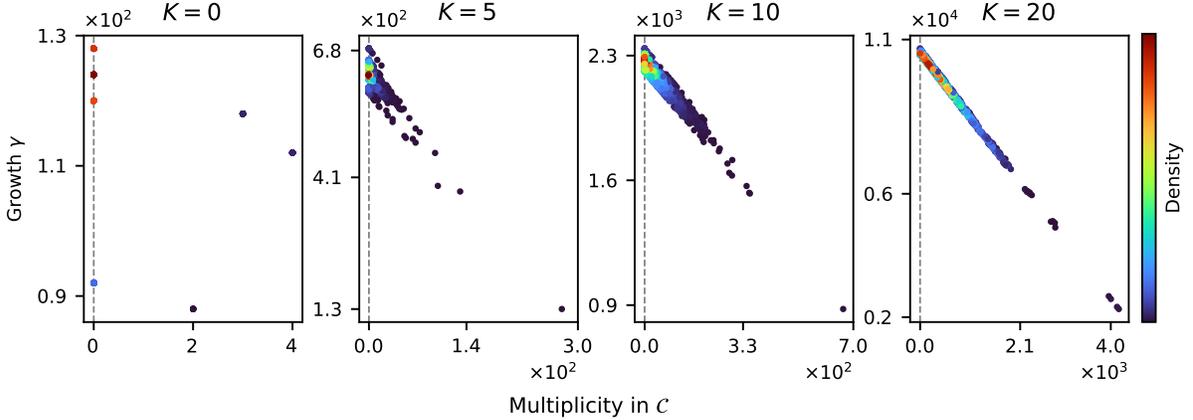}
  \caption{Comparison of multiplicity in $\mathcal{C}$ and growth incurred by Pauli products in the top gradient partition for various effective Hamiltonians along the GM($1$)-iQCC procedure applied to the  CAS($6$e, $6$o)  $N_2$ system at bond length $R=1.5$ \AA{} in the cc-pVDZ basis. $K$ denotes the usage of the $K^{\rm{th}}$ iteration effective Hamiltonian, $\hat H^{(K)}$. For each plot, the multiplicities in $\mathcal{C}$ and growths for all $2^{N-1} = 2048$ Pauli products in the top gradient partition of the Pauli pool for $\hat H^{(K)}$ are shown. Points in regions of greatest density are coloured red, e.g., $\hat H^{(0)}$ only has seven unique data points for all $2048$ Pauli products, with a dominating majority having zero multiplicity in $\mathcal{C}$.}
  \label{fig:n2_corrs}
\end{figure*}

Firstly, we iterate over pairs $(\vec{\mu}^{(i)}, \vec{\mu}^{(j)})$ formed from the set of unique $x$-strings in the Hamiltonian, $\{ \vec{\mu}^{(i)} \}_{i}$, and save pairs which satisfy
\begin{align} \label{eq:x_string_factors}
\left( \vec{\mu}^{(i)} + \vec{\mu}^{(j)} \right) \mod 2 = \vec{\mu}^{(k)}.
\end{align}
In other words, the first step of the algorithm consists of finding factorizations of the $x$-string of interest in terms of pairs of $x$-strings found in the Hamiltonian. In practice one observes that $|\{ \vec{\mu}^{(i)} \}_{i}| \ll M$, and hence this procedure, scaling $O(|\{ \vec{\mu}^{(i)} \}_{i}|^2)$, is not considered relatively burdensome. Once we have the pairs $\{ (\vec{\mu}^{(i)}, \vec{\mu}^{(j)}) \}$ which satisfy Eq. (\ref{eq:x_string_factors}), for each pair $(\vec{\mu}^{(i)}, \vec{\mu}^{(j)})$, we iterate over the terms in $\sum_l c_{l}^{(i)} \hat Z_l^{(i)}$ and $\sum_{l^{'}} c_{l^{'}}^{(j)} \hat Z_{l^{'}}^{(j)}$, i.e. the generalized Ising Hamiltonians multiplying $\hat X_i$ and $\hat X_j$ respectively in Eq. (\ref{eq:ising_H_maintext}). While iterating, we are computing a subset of $\mathcal{C}$ (Eq. (\ref{eq:commutator_set})), the commutator between elements of $\mathcal{H}$ with $x$-strings given by $\vec{\mu}^{(i)}$ and $\vec{\mu}^{(j)}$:
\begin{align} \label{eq:set_C_ij}
    \mathcal{C}_{ij} = \left \{ [Z_{l}^{(i)} \hat X_i, Z_{l^{'}}^{(j)} \hat X_j] \right \}_{l,l^{'}} / \{ 0 \}.
\end{align}
We compute $\mathcal{C}_{ij}$ for all valid $x$-string factorizations $\{ (\vec{\mu}^{(i)}, \vec{\mu}^{(j)}) \}$, and in doing so, obtain sum of multiplicities for any $\hat P_\alpha \in \bigcup_{(ij)} \mathcal{C}_{ij}$ as $\sum_{(ij)} m_{\mathcal{C}_{ij}}(\hat P_\alpha)$. In principle, by first finding valid factorizations of the $x$-string of interest, $\vec{\mu}^{(k)}$, we avoid querying the commutator between Hamiltonian elements which will not possess the correct $x$-string upon commutation, and hence will not generally result in the desired gradient $g^{(k)}$. 
To obtain the growths $\gamma_k$ for $\hat P_\alpha \in \bigcup_{(ij)} \mathcal{C}_{ij}$, we must compute $|\mathcal{H}_A^{(\alpha)}|$ for all considered $\hat P_\alpha$. This step can still be prohibitively expensive for large $M$, even if $|\bigcup_{(ij)} \mathcal{C}_{ij}| \ll |\mathcal{C}|$. Instead of exhaustively computing $|\mathcal{H}_A^{(\alpha)}|$ for all unique $\hat P_\alpha \in \bigcup_{(ij)} \mathcal{C}_{ij}$, we instead only compute $|\mathcal{H}_A^{(\alpha)}|$ for the Pauli products possessing the top $r$ multiplicities. The growth counts are given by $\gamma_\alpha = |\mathcal{H}^{(\alpha)}_A| - 2 \sum_{(ij)} m_{\mathcal{C}_{ij}}(\hat P_\alpha)$. For instance, setting $r=\lceil \log_2 M \rceil$ results in $O(M \log (M))$ scaling for computing $|\mathcal{H}_A^{(\alpha)}|$ for the top $r$ Pauli products by multiplicity. There is generally a strong negative correlation between multiplicities and growths, and hence one is often able to set $r \ll | \bigcup_{(ij)} \mathcal{C}_{ij}| $ while still successfully obtaining the lowest growth count element, with growth $\gamma^{(k)}$. Empirical evidence of the correlation between multiplicities and the growth incurred by Pauli products can be seen in Figure \ref{fig:n2_corrs}, where the multiplicities versus growth for all $2^{N-1}$ Pauli products from the highest $g^{(k)}$ pool partition have been plotted for several effective Hamiltonians for the CAS($6$e, $6$o) N$_2$ model along the GM($1$)-iQCC procedure. Notably, for all considered instances, there are no cases where Pauli product $\hat P_\alpha$ not found in $\mathcal{C}$ (i.e., has zero multiplicity) exhibits lower Hamiltonian growth than all $\hat P_\alpha \in \mathcal{C}$.

\subsubsection{Approximate probabilistic algorithm}

\begin{algorithm}[H]
  \caption{Probabilistic variant of Algorithm \ref{alg:det_alg}.}
  \label{alg:prob_alg}
  \hspace*{\algorithmicindent} \textbf{Input} $\mathcal{H}$, $\vec{\mu}^{(k)}$, $r$, $N_{\rm{samples}}$ \\
    \hspace*{\algorithmicindent} \textbf{Output} smallest growth element $\hat P_\alpha$ with gradient $g_\alpha$ and growth $\gamma^{(k)}$
   \begin{algorithmic}[1]
   \State{obtain set $\boldsymbol{\mu}$}
   \State{initiate $\mathcal{C} = \{ \}$}
   \State{initiate $\texttt{valid\_pairs} = \{\}$}
   \For{$\vec{\mu}^{(i)}, \vec{\mu}^{(j)} \in \boldsymbol{\mu}$, $i < j$}
        \If{$\vec{\mu}^{(i)} + \vec{\mu}^{(j)} \mod 2 = \vec{\mu}^{(k)}$}
            \State{update $\texttt{valid\_pairs} \to \texttt{valid\_pairs} \cup \{ (\vec{\mu}^{(i)}, \vec{\mu}^{(j)}) \}$} 
        \EndIf 
   \EndFor
   \For{sample in $1, 2, \hdots n_{\rm{samples}}$}
        \State{uniformly select $(\vec{\mu}^{(i)}, \vec{\mu}^{(j)}) \in \texttt{valid\_pairs}$}
        \State{uniformly select $\hat Z_l^{(i)} \in \mathcal{Z}_i$, $\hat Z_{l'}^{(j)} \in \mathcal{Z}_j$}
        \State{compute $\hat C = [\hat Z_l^{(i)} \hat X_i, \hat Z_{l^{'}}^{(j)} \hat X_j]$}
        \If{$\hat C \neq 0$}
            \State{update $\mathcal{C} \to \mathcal{C} + \{ \hat C \}$}
        \EndIf
   \EndFor
   \State count occurrences $m_{\mathcal{C}}(\hat P_\alpha)$ for each unique $\hat P_\alpha \in \mathcal{C}$
   \For{$r$ highest occurring $\hat P_\alpha \in \mathcal{C}$}
        \State compute $|\mathcal{H}_A^{(\alpha)}|$
        \State compute and save $\gamma_\alpha = \left | [\hat P_\alpha, \mathcal{H}] / \mathcal{H}_A^{(\alpha)} \right |$
   \EndFor 
   \State select $\hat P_{\alpha^{'}}$ with smallest $\gamma_{\alpha^{'}}$. Then $\gamma^{(k)} = \gamma_{\alpha^{'}}$
   \end{algorithmic}
\end{algorithm}

For problem instances featuring large $M$, Algorithm \ref{alg:det_alg} can be bottlenecked when computing $\mathcal{C}_{ij}$ for all valid $x$-string pairs. To alleviate time and memory requirements of this step, we propose a modified version of Algorithm \ref{alg:det_alg} which utilizes uniform sampling. First, we uniformly sample valid $x$-string pairs, i.e. uniformly select which set we sample from in $\{ \mathcal{C}_{ij} \}_{(ij)}$. Then, instead of computing the selected $\mathcal{C}_{ij}$ set exhaustively, we sample an element of $\mathcal{C}_{ij}$ by uniformly sampling $l$ and $l^{'}$ (see Eq. (\ref{eq:set_C_ij})). The resulting commutator constitutes a ``sample" and is saved if it is non-zero. Since generally growths are negatively correlated with multiplicities in $\bigcup_{(ij)} \mathcal{C}_{ij}$, and higher multiplicity elements are  uniformly sampled from $\bigcup_{(ij)} \mathcal{C}_{ij}$ with higher probability (the probability of sampling a Pauli product $\hat P_\alpha$ is given by $\sum_{(ij)} m_{\mathcal{C}_{ij}}(\hat P_\alpha) / | \sum_{(ij)} \mathcal{C}_{ij}  |$), we find that this probabilistic variant of Algorithm \ref{alg:det_alg} finds the lowest growth element with high success rate and with dramatically less resources. We found empirically that setting the number of samples to be a linear function of $M$ successfully finds the lowest growth element for systems studied in this work. Similarly to Algorithm \ref{alg:det_alg}, we then compute $|\mathcal{H}_A^{(k)}|$ for the $r$ most sampled Pauli products. However, we have not computed all of $\mathcal{C}$, rather we have only sampled elements from it $N_{\rm{samples}}$ times. Hence, the values of $m_{\mathcal{C}} (\hat P_\alpha)$ are not readily available. Thus we compute $\gamma_\alpha$ as $\left | [\hat P_\alpha, \mathcal{H}] / \mathcal{H}_A^{(\alpha)} \right | $. We summarize the probabilistic routine in Algorithm \ref{alg:prob_alg}.

\begin{table}
\caption{A comparison of performance between the deterministic (``Det.") and probabilistic (``Prob.") minimum growth search algorithms of  Algorithm \ref{alg:det_alg} and Algorithm \ref{alg:prob_alg} respectively. For the probabilistic algorithm, $N_{\rm{samples}} = M$ was used. For all examples tabulated, $\gamma^{(k)}$ was obtained for the highest gradient. Observed $\gamma^{(k)}$ values for the probabilistic algorithm are the averages of the lowest growth values computed over ten runs, with standard deviations in brackets. For each effective Hamiltonian, all ten runs find the minimal growth element found in the deterministic algorithm, hence all standard deviations are zero.}
\begin{tabular}{|c|c|c|c|c|}
\hline
\rule{0pt}{2.5ex} Iteration  & $M$ & Algorithm & $N_{\rm{query}}$ & Observed $\gamma^{(k)}$ \\
\hline
\rule{0pt}{2.5ex} \multirow{2}{*}{$0$} & \multirow{2}{*}{$247$} & \multicolumn{1}{c|}{Det.} &  \multicolumn{1}{c|}{$540$} & \multicolumn{1}{c|}{$112$} \\
                                              & & \multicolumn{1}{c|}{Prob.} & \multicolumn{1}{c|}{$247$} & \multicolumn{1}{c|}{$112.0 (0.0)$} \\\cline{1-5}
\rule{0pt}{2.5ex} \multirow{2}{*}{$5$} & \multirow{2}{*}{$1.24 \times 10^3$} & \multicolumn{1}{c|}{Det.} &  \multicolumn{1}{c|}{$1.37 \times 10^{4}$} & \multicolumn{1}{c|}{$130$} \\
                                              & & \multicolumn{1}{c|}{Prob.} & \multicolumn{1}{c|}{$1.24 \times 10^3$} & \multicolumn{1}{c|}{$130.0 (0.0)$} \\\cline{1-5}
\rule{0pt}{2.5ex} \multirow{2}{*}{$10$} & \multirow{2}{*}{$4.47 \times 10^3$} & \multicolumn{1}{c|}{Det.} &  \multicolumn{1}{c|}{$1.29 \times 10^5$} & \multicolumn{1}{c|}{$878$} \\
                                              & & \multicolumn{1}{c|}{Prob.} & \multicolumn{1}{c|}{$4.47 \times 10^3$} & \multicolumn{1}{c|}{$878.0 (0.0)$} \\\cline{1-5}
\rule{0pt}{2.5ex} \multirow{2}{*}{$20$} & \multirow{2}{*}{$2.10 \times 10^4$} & \multicolumn{1}{c|}{Det.} &  \multicolumn{1}{c|}{$1.99 \times 10^6$ } & \multicolumn{1}{c|}{$2263$} \\
                                              & & \multicolumn{1}{c|}{Prob.} & \multicolumn{1}{c|}{$2.10 \times 10^4$} & \multicolumn{1}{c|}{$2263.0 (0.0)$} \\\cline{1-5}
\hline
\end{tabular}
\label{table:det_prob_comparison_n2}
\end{table}

A comparison between the deterministic and probabilistic algorithms is summarized in Table \ref{table:det_prob_comparison_n2}. Both algorithms were applied to the  CAS($6$e, $6$o)  $N_2$ system at bond length $R=1.5$ \AA{} in the cc-pVDZ basis. Effective Hamiltonians at various iterations of the $s(1,0)$ iQCC procedure applied to this system were used for assessment. The number of Hamiltonian Pauli products with coefficient values with magnitude above $10^{-8}$ is denoted by $M$. The main cost of the algorithms is encapsulated by $N_{\rm{query}}$, which refers to the number of times the algorithm must query the commutator function between two Pauli products. Note that non-zero commutators are saved, hence time and memory requirements of both algorithms scale with $N_{\rm{query}}$. For both deterministic and probabilistic algorithms, $r= \lceil \log_2 M \rceil$ was used. We observe that even for large scale effective Hamiltonians, setting $N_{\rm{samples}} = M$ in Algorithm \ref{alg:prob_alg} always finds the lowest growth element found in the deterministic approach of \ref{alg:det_alg}. This is significant, since in this case we query the commutator function between two Pauli products $N_{\rm{query}} = M$ times using Algorithm \ref{alg:prob_alg}, versus $N_{\rm{query}} = O(M^2)$ using Algorithm \ref{alg:det_alg}. In practice, we observe up to $\sim 100\times$ reductions in the number of commutator queries resulting in immensely reduced time and memory requirements, while maintaining identical performance in terms of the lowest observed growth element.

\end{appendix}

\bibliography{main}

\end{document}